\documentclass[aip,jcp,amsmath,amssymb,twocolumn,reprint]{revtex4-1}

\usepackage{amsmath}
\usepackage{amssymb}
\usepackage{amscd}
\usepackage[pdftex]{graphicx}
\usepackage{nicefrac}
\usepackage[np,autolanguage]{numprint}
\usepackage{longtable}
\usepackage{bm}
\usepackage{float}
\usepackage[caption = false]{subfig}

\newcommand{\be}{\begin{equation}}
\newcommand{\ee}{\end{equation}}  
\newcommand{\bea}{\begin{eqnarray}}
\newcommand{\eea}{\end{eqnarray}}

\newcommand{\Fig}{Fig. \ref}
\newcommand{\Tab}{Table. \ref}

\begin{document}

\npthousandsep{}

\title{Binding Energies from Diffusion Monte Carlo for the MB-pol H$_2$O and D$_2$O Dimer: A Comparison to Experimental Values.}

\author{Joel D. Mallory}
\author{Vladimir A. Mandelshtam}
\affiliation{Department of Chemistry, University of California, Irvine, 1102 Natural
Sciences II Irvine, California 92697, USA} 

\begin{abstract}
The Diffusion Monte Carlo (DMC) method is applied to compute the ground state energies of the water monomer and dimer and their D$_2$O isotopomers 
using MB-pol; the most recent and most accurate \textit{ab inito}-based potential energy surface (PES). MB-pol has already demonstrated excellent 
agreement with high level electronic structure data, as well as agreement with some experimental, spectroscopic, and thermodynamic data. Here, the DMC 
binding energies of (H$_2$O)$_2$ and (D$_2$O)$_2$ agree with the corresponding values obtained from velocity map imaging within, respectively, 0.01 
and 0.02 kcal/mol. This work adds two more valuable data points that highlight the accuracy of the MB-pol PES. 
\end{abstract}


\pacs{}
\keywords{}

\maketitle
The development of a full-dimensional potential energy surface (PES) for a many-body system that extends to progressively larger cluster sizes and, at the same 
time, is computationally feasible has been a longstanding issue in electronic structure theory\cite{chalasinski2000}. Many different empirical water PESs have been 
parametrized ranging from fully coarse-grained to flexible atomistic models that include polarization effects and charge transfer (see, e.g., 
Refs.~\citenum{dang1987,fanourgakis2006,fanourgakis2008,burnham2008,molinero2009,habershon2009,vega2011,lee2011}). Nevertheless, empirical PESs are generally 
inadequate for capturing the behavior of water for a wide array of cluster sizes (i.e., from small clusters to the bulk liquid) and thus, have generated ample 
motivation for the construction of \textit{ab initio} and \textit{ab inito}-based surfaces with the latter having a foundation in the many-body expansion of the 
interactions\cite{mayer1940}. Several notable PESs belonging to this family are DPP2\cite{kumar2010}, CC-pol\cite{bukowski2007}, the HBB0-2 
series for the water dimer\cite{huang2006,huang2008,shank2009}, WHBB\cite{wang2011}, and HBB2-pol\cite{babin2012,medders2013}. Along the same vein, the MB-pol 
PES\cite{babin2013_1,babin2014,medders2014} has most recently emerged as an \textit{ab inito}-based water surface rigorously derived from the many-body expansion of 
the interaction energy and expressed in terms of explicit one-, two-, and three-body contributions, with all higher-order terms being represented by (classical) many-body 
induction within a modified version of the polarization model originally employed by the TTM4-F potential\cite{burnham2008}. Similarly to WHBB and HBB2-pol, the MB-pol one-, 
two-, and three-body terms were obtained from fits to large sets of CCSD(T) monomer, dimer, and trimer energies calculated in the complete basis set limit.

Paesani and co-workers (see Refs.~\citenum{babin2013_1,babin2014,medders2014}) have demonstrated that MB-pol effectively reproduces high-level 
\textit{ab initio} data for: (a) stationary point energies on the (H$_2$O)$_2$ and (H$_2$O)$_3$ PESs; (b) PESs for (H$_2$O)$_2$ and (H$_2$O)$_3$ 
interaction energies plotted versus O-O distance and O-O-O angle, respectively; and (c) (H$_2$O)$_4$ interaction energies and (H$_2$O)$_{4-6}$ isomer 
energies. Likewise, results from MB-pol were shown to exhibit good agreement with experimental data for  (H$_2$O)$_2$ vibration-rotation tunneling 
splittings \cite{babin2013_1}, as well as  the structural, thermodynamic and dynamical properties of bulk water at a fully quantum-mechanical level\cite{medders2014}. 
Both infrared and Raman spectra calculated from centroid molecular dynamics simulations with the MB-pol potential were found to be in good agreement with the corresponding 
experimental results.\cite{medders2015_1,medders2015_2}. 

In this paper, we use the diffusion Monte Carlo (DMC) method  
to compute the MB-pol (H$_2$O)$_2$ and (D$_2$O)$_2$ binding energies and compare them to the corresponding experimental values\cite{reisler2011,reisler2012}. 
Note here that Ref.~\citenum{shank2009} reported a DMC result for the binding energy of the HBB2 water dimer, (H$_2$O)$_2$, that was later confirmed experimentally\cite{reisler2011}.

DMC uses a population of $N_W$ random walkers that sample the configuration space bounded by a PES and collectively represent the wavefunction of the many-body system at projection 
time $\tau$ for each time step of length $\Delta\tau$\cite{anderson1975,anderson1976}. At sufficiently long $\tau$ the distribution of random walkers becomes stationary and the 
instantaneous energy $E_{\rm  ref}(\tau)$ fluctuates about its average value $\langle E_{\rm ref}\rangle$. In the limit of $\tau\to\infty$, $\Delta\tau\to 0$, and $N_W\to\infty$, this 
distribution converges to the ground state wavefunction with $\langle E_{\rm ref}\rangle=E_0$, the ground state energy. In a recent paper\cite{mallory2015}, we have undertaken a thorough 
analysis of the behavior and extent of the bias (systematic error) arising from the finite time step $\Delta\tau$, as well as the bias caused by a finite random walker population $N_W$ 
for the q-TIP4P/F\cite{habershon2009} water monomer, dimer and hexamer. The time step bias in the estimate of $E_0$ vanishes slowly, and for the water monomer at $\Delta\tau=10.0$ au is 
$0.015$ kcal/mol. However, this bias cancels nearly completely when the ground state energy differences, such as the binding energy, 
\be
D_0:=2E_{\rm H_2O}-E_{\rm (H_2O)_2},
\ee
or the isotope shift,
\be
\delta D_0:=D_0(\rm D_2O)-D_0(\rm H_2O),
\ee
are computed. (Note that the same value of $\Delta\tau=10.0$ au was used in the DMC calculations for the WHBB water hexamer\cite{wang2012}.) The bias in $N_W$ also cancels for the 
isotope shift ($\delta D_0$), but for the binding energy ($D_0$) it does not cancel. This is because the DMC ground state energy estimate for the monomer converges much faster with 
respect to $N_W$ than that for the dimer. Thus, in order to obtain an accurate estimate of the asymptotic value of $D_0$ at $N_W\to\infty$, one needs to perform a series of calculations
using several different values of $N_W$. It was concluded that $0.002$ kcal/mol accuracy for the dimer binding energy $D_0$ could be achieved using $N_W \sim 2.0\times 10^4$.

In this work, we compute the energies for the MB-pol water monomer and dimer adhering to the DMC protocol of Ref.~\citenum{mallory2015} and performing studies for the time step with 
$\Delta\tau=5$, $10$, and $15$ au and for the random walker population with $N_W=1.6\times 10^3$, $3.6\times 10^3$, and $1.96\times 10^4$. All of the DMC simulations were run to a maximum 
projection time of $2.0\times 10^6$ au. In the time step studies, the walker population was fixed at $N_W=1.96\times 10^4$, while a time step of $\Delta\tau=10.0$ au was used in every 
walker number study. The asymptotic behavior  of the DMC estimate of the ground state energy, $E_0 (\Delta\tau)$, in the $\Delta\tau\to 0$ limit can be approximated well by a polynomial 
with a vanishing derivative at $\Delta\tau =0$. The latter is because the error caused by the split-operator approximation implemented in the DMC method is quadratic in $\Delta\tau$. 
Consequently, since three different values of the time step $\Delta\tau$ have been used, we consider the cubic interpolation:
\be
E_0 (\Delta\tau)\approx A+ B \Delta\tau ^2+ C \Delta\tau ^3,
\ee
using three fitting parameters, with $A$ giving the ground state energy estimate.

Here, we also comment that our walker number studies employ larger values of $N_W$ than might seem to be needed, as including more walkers in the population simultaneously reduces the 
statistical uncertainty as well as the systematic error\cite{mallory2015}.

\begin{figure*}
\caption{DMC energies for the MB-pol water monomer and dimer as a function of the time step $\Delta\tau$. The time steps  
used in this study were $\Delta\tau=5.0$, $10.0$, and $15.0$ au. The walker population was fixed at $N_W=1.96\times 10^4$, 
and the total projection time for all runs was $2.0\times 10^6$ au. Data points were interpolated using a cubic polynomial 
fit with a vanishing derivative at $\Delta\tau\to 0$. Top left: H$_2$O monomer energies. Top right: D$_2$O monomer energies. 
Middle left: H$_2$O dimer energies. Middle right: D$_2$O dimer energies. Bottom left: H$_2$O dimer binding energy. Bottom right: 
D$_2$O dimer binding energy.}
\label{fig:timestep}
\includegraphics[width=0.999\textwidth]{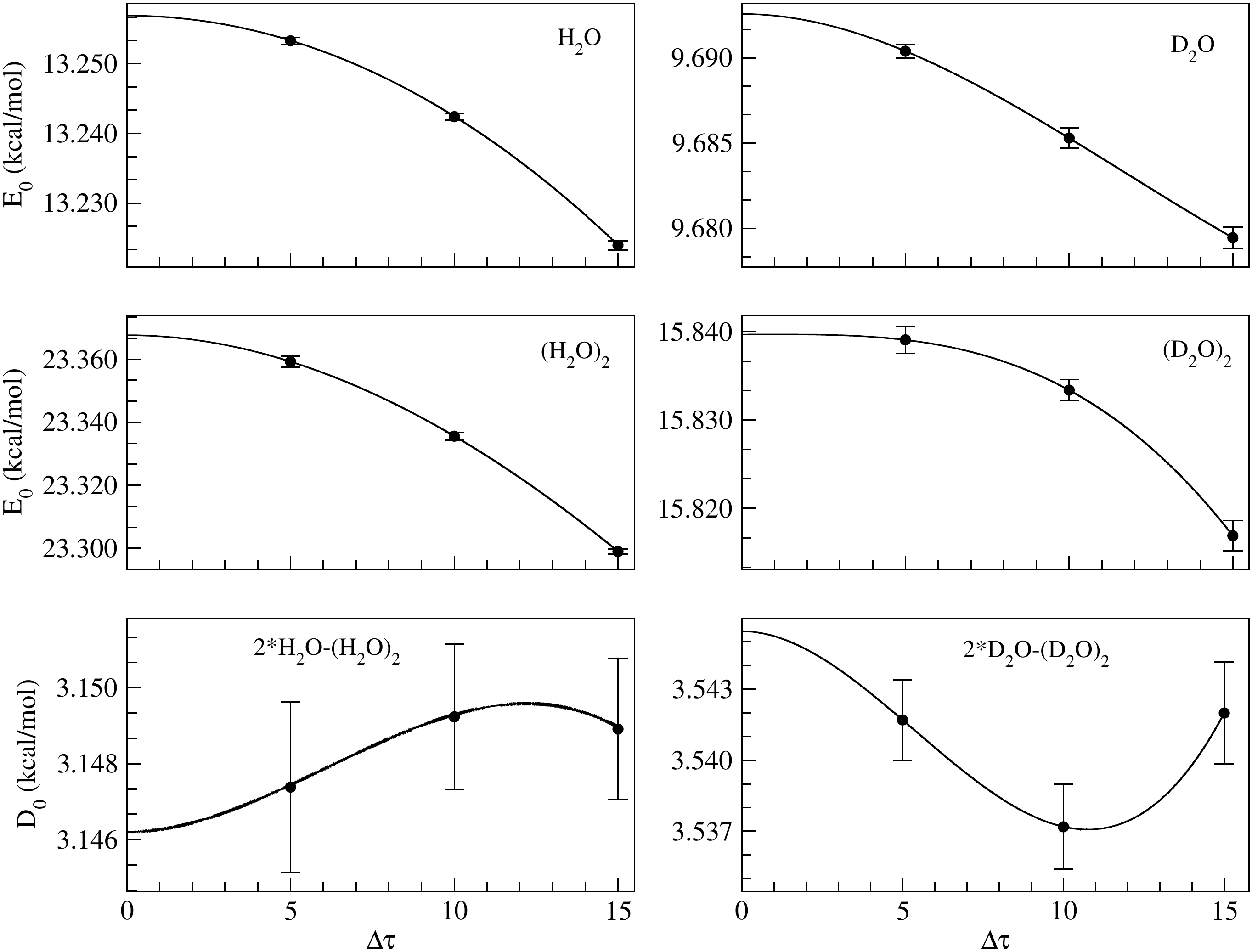}
\end{figure*}

\begin{figure*}
\caption{Same as \Fig{fig:timestep} but as a function of reciprocal walker number $1/N_W$. The walker populations used in this study 
were $N_W=1.6\times 10^3$, $3.6\times 10^3$, and $1.96\times 10^4$. The time step was fixed at $\Delta\tau=10.0$ au, and no special 
interpolation technique was employed.}
\label{fig:nw}
\includegraphics[width=0.999\textwidth]{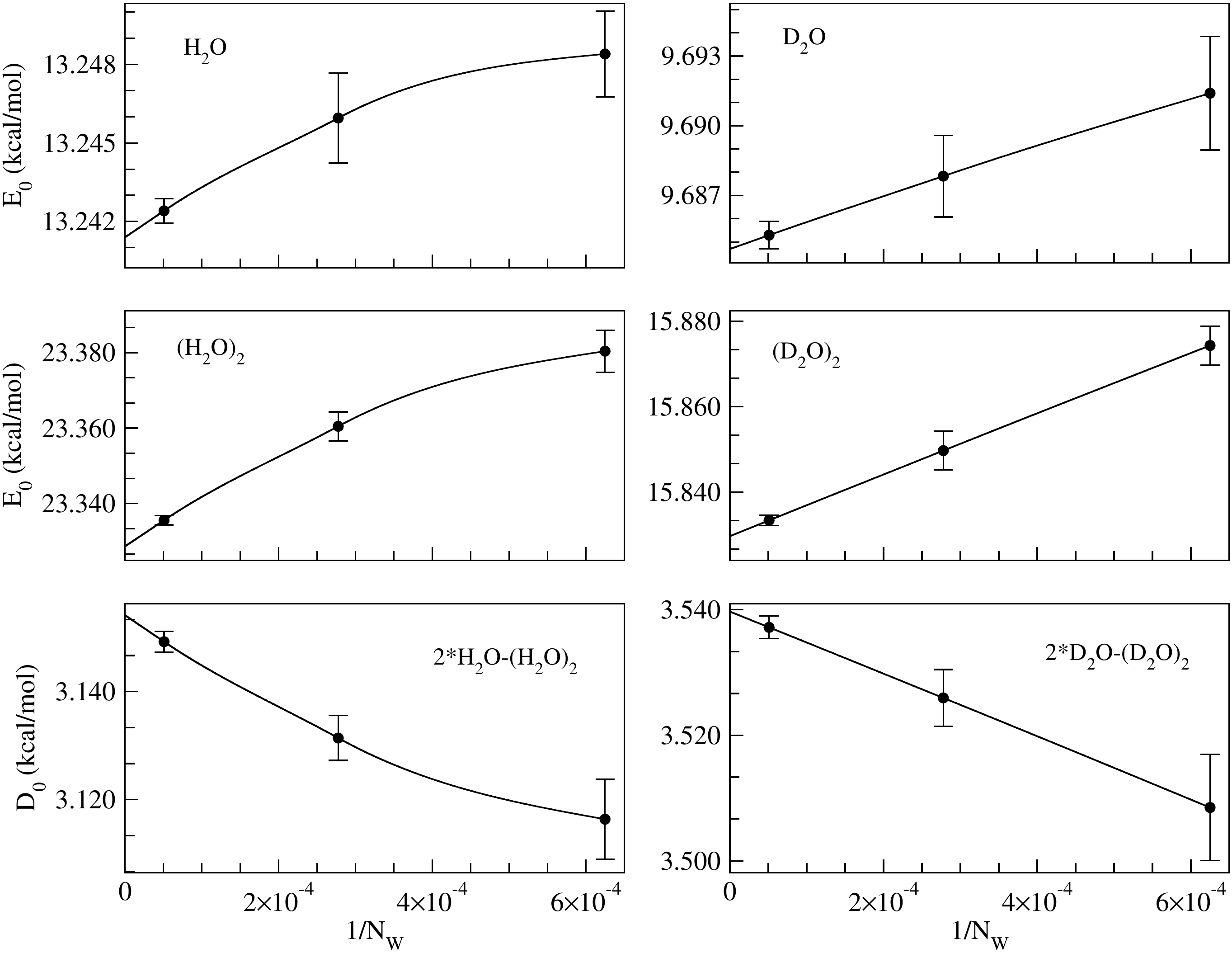}
\end{figure*}

\Fig{fig:timestep} and \Fig{fig:nw} show the bias in $\Delta\tau$ and $N_W$ for the ground state energy estimates from DMC ($E_0$) as functions of the time step and the inverse 
walker population $1/N_W$, respectively, as well as the associated binding energies ($D_0$). The curves follow a pattern similar to that reported in Ref.\citenum{mallory2015} for the q-TIP4P/F 
dimer and monomer. A clear bias does exist for the ground state energies ($E_0$) in both $\Delta\tau$ and $N_W$, but for the reported range of time steps and walker numbers, the energy change 
is in the second or third positions beyond the decimal. Moreover, the binding energy bias in $\Delta\tau$ is virtually negligible due to error cancellations. This is because the largest 
contribution to the time step error comes from the intramolecular degrees of freedom, which are essentially the same regardless of whether the system is comprised of a single water molecule or 
multiple, interacting water molecules. Because the intramolecular modes are nearly invariant to changes in the number and spatial orientation of the water molecules in clusters, the behavior and 
extent of the E$_0$ bias curves in $\Delta\tau$ for the H$_2$O and D$_2$O monomer and dimer are very similar (see the top four panels of \Fig{fig:timestep}), which leads to substantial elimination 
of the bias such that the binding energy displays only a weak dependence on $\Delta\tau$. On the other hand, the bias in $1/N_W$ persists for the binding energy even after the energy differences 
are taken. In this case, the E$_0$ monomer bias in $N_W$ disappears much faster than that of the dimer (note the difference in scales between the monomer (top) and dimer (middle) panels of \Fig{fig:nw}), 
thereby causing imperfect cancellation of error and the appearance of a strong residual bias in the binding energy.  

\begin{table*}
\caption{DMC binding energy $D_0$ for the H$_2$O and D$_2$O dimer with different PESs. The DMC binding energies computed in this work for a time step of $\Delta\tau=10.0$ au were 
extrapolated to the $N_W\to\infty$ limit. Binding energies marked with the ``a'' superscript were reported in ref.\cite{shank2009}. The experimental binding energies (Expt.) are from 
refs.\cite{reisler2011,reisler2012}. The error bar magnitudes for this work are on the order of $10^{-3}$. All energies are reported in kcal/mol.} 

\label{tab:d0}
\begin{tabular}{c|ccccc}
\hline\hline
\multicolumn{6}{c}{$D_0$} \\
\hline
Structure & q-TIP4P/F & TTM3/F & MB-pol & HBB2 & Expt. \\
(H$_2$O)$_2$ & 4.53 & 3.78, $3.84\pm 0.07^a$ & 3.15 & $3.15\pm 0.01^a$ & $3.16\pm 0.03$ \\
(D$_2$O)$_2$ & 4.97 & 4.09 & 3.54 & - & $3.56\pm 0.03$ \\
\hline\hline
\end{tabular}
\end{table*}

\Tab{tab:d0} shows the DMC binding energies extrapolated to the $N_W\to\infty$ limit for three different PESs examined in this work: q-TIP4P/F\cite{habershon2009}, TTM3/F\cite{fanourgakis2008}, 
and MB-pol.\cite{babin2013_1,babin2014,medders2014} The statistical errors for the binding energies with the largest walker population used in this study of $N_W=1.96\times 10^{4}$
are all known to be on the order of $10^{-3}$ kcal/mol, but these numbers are not displayed in the table because extrapolation of the error bar values has proven to be unreliable. Here, the same 
DMC procedure was used to calculate the ground state energies for the TTM3/F H$_2$O and D$_2$O monomer and dimer as those with q-TIP4P/F and MB-pol. Additionally, we also 
provide the (H$_2$O)$_2$ TTM3/F and HBB2 binding energies as reported in Ref.~\citenum{shank2009}. The D$_0$ values for the q-TIP4P/F dimers are $\sim 1.4$ kcal/mol 
larger than the experimental binding energies obtained from velocity map imaging (Expt.)\cite{reisler2011,reisler2012}. This discrepancy is not surprising considering that the q-TIP4P/F 
PES was not designed to represent the microscopic properties of small water clusters accurately. Our TTM3-F D$_0$ value for (H$_2$O)$_2$ agrees well with that reported by Bowman and 
co-workers\cite{shank2009} with a difference of only $0.06$ kcal/mol, but our results indicate that the binding energy for this PES is still off from the experimental values by $0.62$ and 
$0.53$ kcal/mol for (H$_2$O)$_2$ and (D$_2$O)$_2$, respectively.

Notably, the (H$_2$O)$_2$ DMC binding energy, also from the \textit{ab initio}-based two-body PES, HBB2\cite{shank2009}, (with values of $\Delta\tau$ and $N_W$ similar to those used in the 
present work) is highly accurate upon comparison to the experimental results. At the same time, we highlight that the MB-pol PES likewise shows nearly perfect agreement with the velocity 
map imaging D$_0$ values for both MB-pol dimers, (H$_2$O)$_2$ and (D$_2$O)$_2$. For the MB-pol PES, the binding energies differ from the experimental results by only $0.01$ kcal/mol for 
(H$_2$O)$_2$ (as is also the case for the HBB2 dimer PES) and $0.02$ kcal/mol for (D$_2$O)$_2$. Such an excellent correspondence adds two more valuable data points which underscore a 
favorable assessment for the accuracy of the MB-pol PES. 

Therefore, efforts are currently underway to establish the true ground state energy and wavefunction of the MB-pol water hexamer using DMC. Additionally, in accordance with 
Ref.\citenum{mallory2015}, the ground state energies for the isomers of the MB-pol hexamer will be computed.

\section*{Acknowledgements}
This work was supported by the National Science Foundation (NSF) Grant No. \mbox{CHE-1152845}.
We thank Sandra Brown for useful discussions, and Francesco Paesani and his co-workers for sharing the
MB-pol code with us.

%

\end{document}